\documentclass[twocolumn,letter]{jpsj2} %% two-column layout
\title{Evidence for Point Nodes in the Superconducting Gap Function in the Filled Skutterudite Heavy-Fermion Compound PrOs$_4$Sb$_{12}$: $^{123}$Sb-NQR Study under Pressure}

\author{Kouta \textsc{Katayama}, Shinji \textsc{Kawasaki}, Masahide \textsc{Nishiyama}, Hitoshi \textsc{Sugawara}$^{1}$, Daisuke \textsc{Kikuchi}$^{2}$, Hideyuki \textsc{Sato}$^{2}$ and Guo-qing \textsc{Zheng}}

\inst{Department of Physics, Faculty of Science, Okayama University, Okayama 700-8530, Japan \\
$^{1}$Department of Mathematical and Natural Sciences, Faculty of Integrated Arts and Sciences,
The University of Tokushima, Tokushima 770-8502, Japan\\
$^{2}$Department of Physics, Graduate School of Science, Tokyo Metropolitan University,
Hachioji, Tokyo 192-0397, Japan}

\recdate{\today}

\abst{We report $^{123}$Sb nuclear quadrupole resonance (NQR) measurements of the filled skutterudite heavy-fermion superconductor PrOs$_4$Sb$_{12}$ under high pressures of  1.91 and 2.34 GPa. The temperature dependence of NQR frequency and the spin-lattice relaxation rate $1/T_1$ indicate that the crystal-electric-field splitting $\Delta_{\rm CEF}$ between the ground state $\Gamma_1$ singlet and the first excited state $\Gamma_4^{(2)}$ triplet decreases with increasing pressure. The 1/$T_1$ below $T_c$ = 1.55 K at $P$ = 1.91 GPa shows a power-law temperature variation and is proportional to $T^5$ at temperatures considerably below $T_c$, which indicates the existence of point nodes in the  superconducting gap function. The data can be well fitted by the gap model $\Delta(\theta)=\Delta_0\sin\theta$ with $\Delta_0$ = 3.08$k_{\rm B}T_{\rm c}$. The  relation between the superconductivity and the quadrupole fluctuations associated with the $\Gamma_4^{(2)}$ state is discussed.
}

\kword{PrOs$_4$Sb$_{12}$, superconductivity, NQR, pressure}

\newpage
\begin{document}
\maketitle

The filled skutterudite compound PrOs$_4$Sb$_{12}$ is the first  praseodymium (Pr)-based heavy-fermion superconductor with $T_c$ = 1.85 K \cite{Bauer}. The heavy-electron mass has been confirmed by the large specific heat jump $\Delta$C/$T_c$ $\sim$ 500 mJ/($K^2$mol) at $T_c$\cite{Bauer,Maple} and by de Haas-van Alphen effect  measurements\cite{Sugawara}. The ground state of the crystal electric field (CEF) for a Pr$^{3+}$ ion is a $\Gamma_1$ singlet, which is separated by the first excited state of the $\Gamma_4^{(2)}$ triplet by a gap of $\Delta_{\rm CEF}$ $\sim$ 10 K.\cite{AokiJPSJ,Tenya,Tayama2003,Kohgi,Kuwahara,Goremychkin} 
Because of this small $\Delta_{\rm CEF}$, the relation between the quadrupole fluctuations associated with the $\Gamma_4^{(2)}$ state \cite{Tenya,Bauer2} and the occurrence of the superconductivity has been the focus of discussions \cite{Koga,Goremychkin,Kuwahara2}.

Although the superconducting gap function is important for understanding the superconductivity in PrOs$_4$Sb$_{12}$, it has not yet been determined experimentally. Previous nuclear-quadrupole-resonance (NQR) measurement  has revealed the unconventional nature of the superconductivity\cite{Kotegawa}.   The spin-lattice relaxation rate 1/$T_1$ shows no coherence peak just below $T_c$, nonetheless, it follows an exponential temperature dependence at low temperatures. On the basis of  this result and the exponential decrease in the penetration depth below $T_c$ found by muon spin relaxation ($\mu$SR)\cite{MacLaughlin}, it has been proposed that the superconducting gap for PrOs$_4$Sb$_{12}$ is isotropic\cite{Kotegawa,MacLaughlin}. In contrast, oscillations with respect to the magnetic field angle have been found in angle-resolved thermal conductivity measurements, which suggest that the superconducting gap is anisotropic, and  the existence of point nodes in the gap function has been proposed \cite{Izawa}. The results of penetration depth measurement also suggested a point-nodes gap \cite{Chia}.  Our previous NQR study on the substitution system Pr(Os$_{1-x}$Ru$_x$)$_4$Sb$_{12}$ strongly suggested the existence of nodes in the superconducting gap function of PrOs$_4$Sb$_{12}$, since Ru doping at the Os site in PrOs$_4$Sb$_{12}$ as a nonmagnetic impurity induces a residual density of states in the superconducting gap\cite{Nishiyama}, but the nodal structure remains unclear.

In this paper, we report on the $^{123}$Sb-NQR study of PrOs$_4$Sb$_{12}$ under pressure. Applying pressure reduces $T_c$ \cite{Maple}  and may also change  $\Delta_{\rm CEF}$ \cite{Tayama}, and therefore can provide new information on the  symmetry of the superconducting gap as well as on the mechanism of the superconductivity.  We find that $\Delta_{\rm CEF}$ decreases with increasing pressure, and that at $P$ = 1.91 GPa, $1/T_1$ decreases  in proportion to $T^5$ at low temperatures, which indicates point nodes in the superconducting gap function.

 Single crystals of PrOs$_4$Sb$_{12}$ were grown by the Sb-flux method. For NQR measurements, the single
crystals were powdered to allow the rf magnetic field to penetrate into the sample. However, the size of the grains is kept larger than 100 $\mu$m to avoid crystal distortions. Hydrostatic pressure was applied using a NiCrAl/BeCu piston-cylinder  cell, filled with Si oil as a pressure-transmitting medium\cite{Andrei}. The pressure at low temperatures was determined from the pressure dependence of the $T_c$ value of Sn metal measured by a conventional four-terminal method. Data below 1.4 K were collected using a $^3$He/$^4$He dilution refrigerator at $P$ = 0 and a $^3$He refrigerator  under high pressure $P$ = 1.91 GPa. To avoid possible heating due to the rf pulses, we used very small amplitude rf pulses in the $T_1$ measurements at low temperatures. We confirmed the lack of such a heating effect  by ensuring that the spin echo intensity is not affected by an rf pulse with a slightly off-resonance frequency, which was applied before the $\pi/2-\pi$ pulse sequence.

%fig1
\begin{figure}[h]
\centering
\includegraphics[width=7cm]{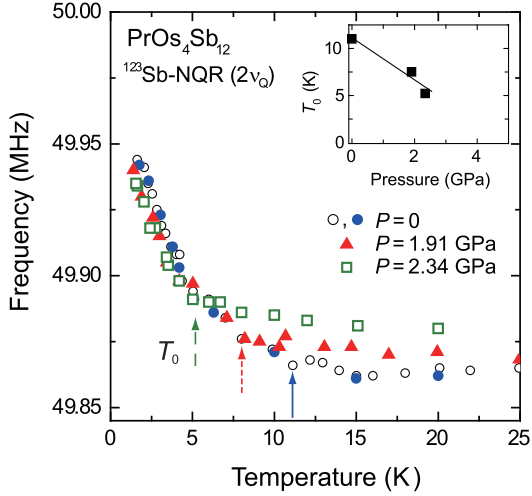}
\caption[]{\footnotesize  (color online) Temperature dependence of 2$\nu_Q$ position of $^{123}$Sb-NQR ($\pm 3/2\leftrightarrow \pm 5/2$ transition) at $P$ = 0 (solid circles), 1.91 (solid triangles), and 2.34 GPa (open squares) along with the data at $P$ = 0 (open circles) by Kotegawa {\it et al} \cite{Kotegawa}. Solid, dotted, and dashed arrows indicate $T_0$ at $P$ = 0, 1.91, and 2.34 GPa, respectively (see text).   
The inset shows the pressure dependence of $T_0$. } 
\end{figure}

 The Sb nuclei have two isotopes of $^{121}$Sb and $^{123}$Sb with natural abundances of 57.3\% and 42.7\%, respectively. 
Since $^{121}$Sb and $^{123}$Sb have the nuclear spins $I$ = 5/2 and 7/2, respectively, five Sb-NQR transitions  are observed.\cite{Kotegawa}  In the present experiment, all measurements were carried out at the $\pm$3/2$\leftrightarrow$ $\pm$5/2 transition (hereafter, 2$\nu_Q$ transition for short) of the $^{123}$Sb nucleus. 
Figure 1 shows the increase in the 2$\nu_Q$ resonance frequency below $T$ = 25 K for various pressures. $T_0$ is the temperature at which the 2$\nu_Q$ resonance frequency increases abruptly. Since the electrical field gradient (EFG) is predominantly determined by the on-site charge distribution, the NQR frequency is a good measure of the population of the ground/excited state. Indeed, in both PrOs$_4$Sb$_{12}$ \cite{Kotegawa} and PrRu$_4$Sb$_{12}$ \cite{Yogi}, $T_0$ is in good agreement with $\Delta_{\rm CEF}/k_B$. More recently,  it has been suggested that the  temperature dependence of NQR frequency can be accounted for  by the EFG associated with the hexadecapole moment of the $\Gamma_4^{(2)}$ state \cite{Tou}. 
Therefore, the increase in the NQR frequency below $T_0$ indicates that the depopulation of the $\Gamma_4^{(2)}$ state occurs below this temperature.
 As seen in Fig. 1,  $T_0$ shifts to lower temperatures $T_0(P)$ $\sim$ 7.5 and 5 K at $P$ = 1.91 and 2.34 GPa, respectively  (also see Fig. 1 inset). These results provide evidence that $\Delta_{\rm CEF}$ decreases with increasing pressure.

 %fig2
\begin{figure}[h]
\centering
\includegraphics[width=7.5cm]{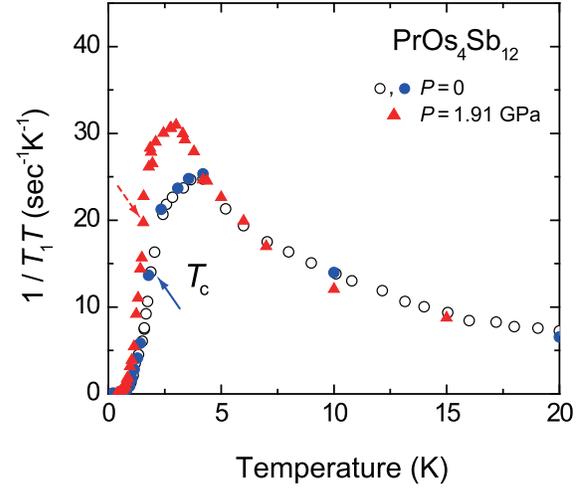}
\caption[]{\footnotesize (color online) Temperature dependence of $1/T_1T$ at $P$ = 0 (solid circles) and 1.91 GPa (solid triangles) along with the data at ambient $P$ cited from literature\cite{Kotegawa} (open circles). Solid and dotted arrows indicate $T_c(P)$ at $P$ = 0 and 1.91 GPa, respectively.}
\end{figure}

The above conclusion is supported by the pressure effect on the temperature dependence of $1/T_1T$. Figure 2 shows the temperature dependence of $1/T_1T$ at $P$ = 0 and 1.91 GPa. The pressure effect appears below 4 K. At $P$ = 0, the reduction of $1/T_1T$  results in a peak in the plot of $1/T_1T$ versus $T$, which is due to the depopulation of the $\Gamma_4^{(2)}$ state below $T_0$.  At $P$ =1.91 GPa, the decrease in $1/T_1T$ occurs at a lower temperature, indicating the decrease in $\Delta_{\rm CEF}$. 
  These results are consistent with the conclusion inferred from the magnetization measurement.\cite{Tayama}  
  
%fig3
\begin{figure}[h]
\centering
\includegraphics[width=7cm]{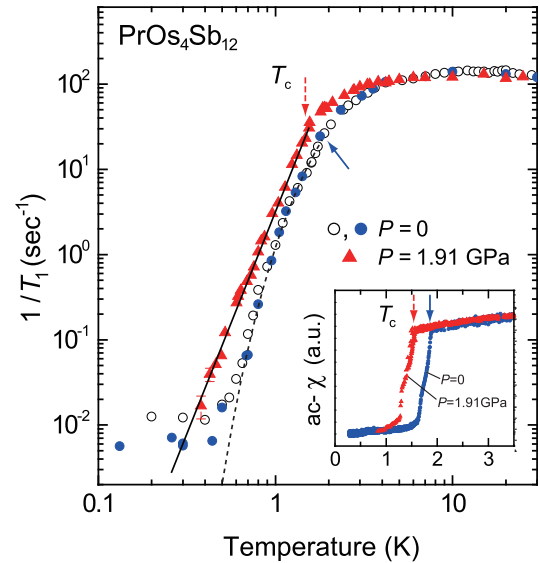}
\caption[]{\footnotesize (color online) Temperature dependence of $1/T_1$ at $P$ = 0 (solid circles) and 1.91 GPa (solid triangles) along with the data at ambient $P$ cited from ref. 13 (open circles). The straight line is a guide to the eyes. The dotted curve depicts the relation $1/T_1$ $\propto$ exp(-$\Delta_0$/$k_{B}T$) with $\Delta_0$/$k_{\rm B}T_{\rm c}$ = 3.45, proposed by Yogi {\it et al} \cite{Yogi2}.
The inset shows the temperature dependence of ac-susceptibility at $P$ = 0 and 1.91 GPa. Solid and dotted arrows indicate $T_c$  at $P$ = 0 and 1.91 GPa, respectively. }

\end{figure}

 We find concomitantly that the temperature of the onset of the superconducting transition decreased with increasing pressure, in agreement with previous reports.\cite{Maple,Tayama} The inset in Fig. 3 shows the temperature dependence of ac-susceptibility measured  using the NQR coil.  $T_c$ decreased from 1.85 K at $P$ = 0 to 1.55 K  at $P$ = 1.91 GPa. 
The main panel of Fig. 3 shows the temperature dependence of  $1/T_1$ at $P$ = 0 and 1.91 GPa. The ambient-pressure data are in excellent agreement with those reported previously (open circles) \cite{Kotegawa}, except that $1/T_1$  below $T\sim$ 0.4 K is smaller in the present sample, probably due to the improvement of the  sample quality \cite{note}.  
The data could be fitted by an exponential function $1/T_1$ $\propto$ exp(-$\Delta_0$/$k_{B}T$) below $T_c$, as pointed out previously \cite{Kotegawa}.
 At both $P$ = 0 and 1.91 GPa, 1/$T_1$ is $T$-independent above $T_0$, indicating that the relaxation in the high-temperature region is predominated by the Pr-4$f^2$-derived localized magnetic moments.  With decreasing temperature below $T_0$, 1/$T_1$ starts to decrease. Below $T_c$, no coherence peak  is observed at $T_c$ = 1.55 K for $P$ = 1.91 GPa, as for $P$ = 0.\cite{Kotegawa}
However,  the $T$ dependence of 1/$T_1$ at high pressure is markedly different from that at ambient pressure. 1/$T_1$ at $P$ = 1.91 GPa decreases in a power law of $T$ below $T_c$.

In particular,  below $T\sim$ 0.55 K, 1/$T_1$ is proportional to $T^5$, as can be seen more clearly in Fig. 4. 
We find that a point-nodes model, with a low-energy ($E$) superconducting density of states (DOS) proportional to  $E^2$, can well explain the experimental result.
In Fig. 4, the curve below $T_c$ is a fit to the  Anderson-Brinkman-Morel (ABM) model  \cite{ABM1,ABM2}. Namely, 

\[
\frac{T_1(T_{\rm c})}{T_{1}}=\frac{2}{k_{\rm B}T} \int \left( \frac{N_{\rm S}(E)}{N_0} \right)^2 f(E) [1-f(E)] dE,
\]
where $N_{\rm S}(E)/N_0=E/\sqrt{E^2-\Delta^2}$  with $\Delta(\theta)=\Delta_0\sin\theta$.
 The fit gives $\Delta_0/k_{\rm B}T_{\rm c}$ = 3.08.
 The penetration depth data at ambient pressure seem to be consistent with our results \cite{Chia}. However, 
  the  $p+h$ models  \cite{Maki} proposed to explain the thermal conductivity would give a $T^3$-like dependence, since the  DOS at low-$E$ is linear in $E$,   and is therefore not compatible with our data.

%fig4
\begin{figure}[h]
\centering
\includegraphics[width=6.5cm]{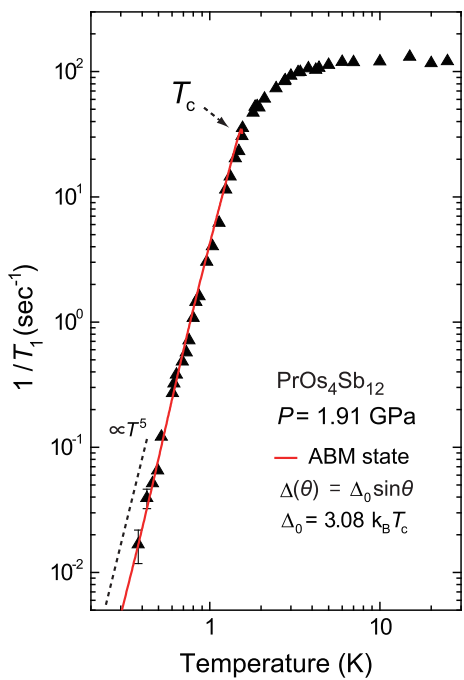}
\caption[]{\footnotesize (color online) Temperature dependence of $1/T_1$ at $P$ = 1.91 GPa. The arrow indicates $T_c$. The solid  curve is a   fit  assuming the ABM state  with  $\Delta_0$/$k_{\rm B}T_{\rm c}$ = 3.08. The dotted line indicates the relation of 1/$T_1$ $\propto$ $T^5$ considerably below $T_c$. }

\end{figure}

  It has been proposed that the superconductivity is mediated by the excitons due to the  $\Gamma_4^{(2)}-\Gamma_1$ quasi-quartet \cite{Koga}. In such case, $T_c$ would increase when $\Delta_{\rm CEF}$ is reduced. Clearly, our results do not lend a straightforward support to this theory. Further experimental study under higher pressure is highly desirable. 
Finally, we comment on the  different temperature dependences of $1/T_1$ at $P$ = 0 and 1.91 GPa. 
Two possible causes could be responsible.  First, the larger gap $\Delta_{\rm CEF}$ may contribute to the reduction of 1/$T_1$ below  $T_c$ at ambient pressure, which makes the temperature dependence of 1/$T_1$ exponential. Second, it may be due to the multiple-band nature of the superconductivity \cite{Flouquet}. Recent thermal conductivity measurement under a magnetic field suggests the superconductivity  at ambient pressure is induced in two different Fermi sheets \cite{Flouquet}, which may have different symmetry. The sheet in which nodes develop may grow significantly under high pressures.

In conclusion, we have presented the $^{123}$Sb-NQR results on the filled skutterudite heavy-fermion compound PrOs$_4$Sb$_{12}$ at $P$ = 0, 1.91, and 2.34 GPa. The temperature dependence of NQR frequency and the spin-lattice relaxation rate $1/T_1$ indicate that the gap $\Delta_{\rm CEF}$ between the ground state $\Gamma_1$ singlet and the first excited state $\Gamma_4^{(2)}$ triplet decreases with increasing pressure. At $P$ = 1.91 GPa, the temperature dependence of 1/$T_1$ below $T_c$ is well explained  by the ABM superconducting state, with point nodes in the gap function.  To confirm the mechanism showing why $T_c$ decreases with increasing pressure in PrOs$_4$Sb$_{12}$, further NQR measurements under pressure are now in progress.

We thank K. Matano for his experimental help, and  R. Shiina, H. Harima, Y. Kitaoka, H. Tou, M. Ichioka, N. Nakai, H. Kotegawa, and  M. Yogi for discussions  and  comments. This work was supported in part by grants for scientific research, and a grant for Attractive Education in Graduate School "Training Program for Pioneers of Frontier and Fundamental Sciences", from the Ministry of Education, Culture, Sports, Science and Technology.

\end{document}